\documentclass[twocolumn,twocolappendix]{aastex631}
\shortauthors{Chen et al.}
\shorttitle{}

\usepackage{subfigure} 

\begin{document}
%\title{The CGM distribution around galaxies traced by Mg\,\textsc{ii} absorption of background QSO: dependence on galaxy properties}

%\title{The circumgalactic medium of galaxies traced by Mg\,\textsc{ii} absorption with DESI: dependence on galaxy properties}

%\title{Cosmic Evolution of the Multiphase Gas Distributions associated with QSOs}

\title{The Cosmic Evolution and Spatial Distribution of Multiphase Gas associated with QSOs}

\author[0009-0004-5989-6005]{Zeyu Chen}\thanks{Email:  \href{mailto:czy664@mail.ustc.edu.cn}{czy664@mail.ustc.edu.cn}}
\affiliation{Department of Astronomy, University of Science and Technology of China, Hefei 230026, China}
\affiliation{School of Astronomy and Space Science, University of Science and Technology of China, Hefei 230026, China}

\author[0000-0003-1588-9394]{Enci Wang}\thanks{Email:  \href{mailto:ecwang16@ustc.edu.cn}{ecwang16@ustc.edu.cn}}
\affiliation{Department of Astronomy, University of Science and Technology of China, Hefei 230026, China}
\affiliation{School of Astronomy and Space Science, University of Science and Technology of China, Hefei 230026, China}

\author[0000-0002-6684-3997]{Hu Zou}\thanks{Email: \href{mailto:zouhu@nao.cas.cn}{zouhu@nao.cas.cn}}
\affiliation{National Astronomical Observatories, Chinese Academy of Sciences, Beijing 100012, China}

\author[0009-0008-1319-498X]{Haoran Yu}
\affiliation{Department of Astronomy, University of Science and Technology of China, Hefei 230026, China}
\affiliation{School of Astronomy and Space Science, University of Science and Technology of China, Hefei 230026, China}

\author[0000-0003-3667-1060]{Zhicheng He}
\affiliation{Department of Astronomy, University of Science and Technology of China, Hefei 230026, China}
\affiliation{School of Astronomy and Space Science, University of Science and Technology of China, Hefei 230026, China}

\author[0000-0002-4911-6990]{Huiyuan Wang}
\affiliation{Department of Astronomy, University of Science and Technology of China, Hefei 230026, China}
\affiliation{School of Astronomy and Space Science, University of Science and Technology of China, Hefei 230026, China}

\author[0000-0002-2851-886X]{Yang Gao}
\affiliation{College of Physics and Electronic Information, Dezhou University, Dezhou 253023, China}
%\affiliation{National Natural Science Foundation of China (NSFC, Nos.12033004 and 12233005)}

\author[0009-0000-7307-6362]{Cheqiu Lyu}
\affiliation{Department of Astronomy, University of Science and Technology of China, Hefei 230026, China}
\affiliation{School of Astronomy and Space Science, University of Science and Technology of China, Hefei 230026, China}

\author[0009-0004-7042-4172]{Cheng Jia}
\affiliation{Department of Astronomy, University of Science and Technology of China, Hefei 230026, China}
\affiliation{School of Astronomy and Space Science, University of Science and Technology of China, Hefei 230026, China}

%\author[0009-0009-2660-1764]{Haixin Li}
%\affiliation{Department of Astronomy, University of Science and Technology of China, Hefei 230026, China}
%\affiliation{School of Astronomy and Space Science, University of Science and Technology of China, Hefei 230026, China}

\author[0009-0006-7343-8013]{Chengyu Ma}
\affiliation{Department of Astronomy, University of Science and Technology of China, Hefei 230026, China}
\affiliation{School of Astronomy and Space Science, University of Science and Technology of China, Hefei 230026, China}

\author[0000-0003-0230-4596]{Weiyu Ding}
\affiliation{Department of Astronomy, University of Science and Technology of China, Hefei 230026, China}
\affiliation{School of Astronomy and Space Science, University of Science and Technology of China, Hefei 230026, China}
\affiliation{National Astronomical Observatories, Chinese Academy of Sciences, Beijing 100012, China}

\author[0000-0002-8037-3573]{Runyu Zhu}
\affiliation{National Astronomical Observatories, Chinese Academy of Sciences, Beijing 100012, China}
\affiliation{Department of Astronomy, Xiamen University, Xiamen, Fujian 361005, China}

\author[0000-0002-7660-2273]{Xu Kong}
\affiliation{Department of Astronomy, University of Science and Technology of China, Hefei 230026, China}
\affiliation{School of Astronomy and Space Science, University of Science and Technology of China, Hefei 230026, China}

%\author{}



\begin{abstract}

\noindent {We investigate the multi-phase gas surrounding QSOs traced by 33 absorption lines (e.g., Ly$\alpha$, C\,\textsc{iv}, Fe\,\textsc{ii}, Mg\,\textsc{ii}, etc.) in the stacked spectra of background sources, using the early data release from the Dark Energy Spectroscopic Instrument. Our analysis reveals that the equivalent width (\( W \)) of metal absorption lines decreases with increasing redshift, following an overall trend described by $W \propto (1+z)^{-4.0\pm 2.7}$. Different species that trace multi-phases of QSO-associated gas exhibit distinct evolutionary patterns.
Additionally, the \( W \) of these absorption lines decreases with distance ($D$) from QSOs, which can be effectively characterized by a two-halo model. Compared to the projected two point correlation function of galaxies at similar redshifts, low-ionization ions exhibit similar clustering scales, while high-ionization ions show a significantly more extended spatial distribution. We also find that $W_{\text{FeII}}/W_{\text{MgII}}$ increases towards lower redshifts, which can be attributed to evolving star formation histories and/or changes in initial mass function for galaxies. By leveraging multiple absorption tracers, we conduct the first comprehensive investigation of diffuse, multiphase gas from the circumgalactic medium to cosmological scales, offering new insights into baryon cycles and the transport of metals throughout cosmic time.
}

\end{abstract}

\keywords{ quasars: absorption lines, galaxies: halos, circumgalactic medium, intergalactic medium}


\section{Introduction} \label{sec:intro}
The absorption lines observed in the spectra of distant quasi-stellar objects (QSOs) \citep{rauch98,berk01,Wolfe05} provide one of the best ways of understanding the evolving chemical enrichment of the Universe \citep[e.g.,][]{dey2015,lan17,lan20,wu2024}. These lines arise in the gas surrounding galaxies along the lines of sight to QSOs, offering valuable insights into the density, kinematics, and chemical composition of the local gas \citep{bahcall1969,yanny1990,bergeron1991,steidel1995,zhu14}. For over four decades, absorption line spectroscopy has been employed to probe the distribution of gas around galaxies, particularly the Circumgalactic Medium \citep[CGM; e.g.,][]{steidel10,zhu13ca,lan18,Zheng24,chen25}. The CGM serves as a dynamic interface between galaxies and the cosmic web, playing a crucial role in the formation and evolution of galaxies \citep{tumlinson17,Wang-19,Wang-22, Wang-23, Michele24, Ma-24, Lyu-25}. It also provides critical insights into both local and large-scale processes that shape the universe \citep{Peeples20,peroux2020, Ayromlou-23, peroux2024}. %Consequently, understanding the origin, distribution, and dynamical evolution of elements in the CGM has emerged as a cornerstone for deciphering galaxy-environment coevolution.

Metal-enriched absorption is widely believed to result from large-scale galactic outflows driven by stellar winds, supernovae (SNe), active galactic nuclei (AGN), and the accretion of gas tidally stripped during mergers from star-forming galactic discs \citep[e.g.,][]{oppenheimer2006,dalla2008,suresh2015}. Compared to massive galaxies, dwarf galaxies offer more opportunities for metals to be transferred to the CGM and even the Intergalactic Medium (IGM) through these outflows \citep{Muratov17, Ayromlou-23}. Observational evidence for such outflows is well-established through the detection of blueshifted Mg\,\textsc{ii} and Fe\,\textsc{ii} absorption lines observed down-the-barrel in galactic spectra, which trace kinematically disturbed gas moving at velocities of 100 to 1000 km/s
\citep[e.g.][]{weiner09,erb2012,kornei2012,martin12,rubin14}. These feedback mechanisms facilitate the bulk transport of metals from disk to halo scales and establish a dynamic cycling process, redistributing enriched material throughout the CGM and the cosmic web \citep{bordoloi11,prochaska11b,rubin14}.

Despite the ability of outflow mechanisms to explain the origin of metals in the gas surrounding galaxies  \citep{shin21}, substantial uncertainties remain regarding the spatial distribution \citep[e.g.,][]{peroux2020,weng24} and temporal evolution \citep[e.g.,][]{dey2015,lan20} of these metals once expelled from the galactic disk. This process involves complex interactions within the multiphase medium, such as outflow deceleration, gas mixing, reaccretion, and interactions with cosmic filaments \citep{tumlinson17,faucher2023,saeedzadeh23}. The retention time, diffusion extent, and chemical state of metals in the gas surrounding galaxies are heavily influenced by the physical conditions of the transport paths they transverse \citep{Maiolino19,Michele24}. 

% The {\tt Eris} simulation indicates that hot metals (\(> 10^5 \, \rm{K} \)) preferentially propagate into the CGM via high-velocity outflows, establishing a radially declining temperature gradient \citep{shen12}. Cool photoionized gas (\(  T \sim 10^4 \, \rm{K} \)) is primarily traced by the C\,\textsc{ii} and Si\,\textsc{ii} absorption lines, which exhibit a sharply decreasing covering factor beyond \( R_{\text{vir}} \). In contrast, high-ionization species such as O\,\textsc{vi} \citep[e.g.,][]{sanchez19,li21} and C\,\textsc{iv} maintain a substantial covering factor beyond \( 2 R_{\text{vir}} \), with O\,\textsc{vi}-enriched halos extending to \( \sim 4 R_{\mathrm{vir}} \) \citep{shen13}. These results align with  observations of star-forming galaxies at low redshift by  \cite{tumlinson11} and  \cite{prochaska11}. 

%\textbf{Simulations, such as \texttt{FIRE}, indicate that turbulence and repeated recycling--re-ejection processes would rapidly mix the metals injected into the inner CGM, while some metal-poor gas is expelled into the IGM \citep{hafen20}. \cite{nishigaki25} recently present a model of metallicities in multiple gas phases from $z = 0$ to 5 and find that, within $z \lesssim 2$, the fallback rate of outflowing metals increases, leading to a rapid growth of metallicity in the CGM, while the IGM remains consistently diffuse and metal-poor.}

Several observations suggest that the distribution of gas around galaxies follows a two-halo model \citep{zhu14,wu2024},  varying across different ions. For instance, \cite{Zahedy17} investigated the chemical composition of cool gas around 27 intermediate-redshift galaxies and found a declining [Fe/Mg] ratio with increasing galactocentric distance. Additionally, a decreasing $W_{\text{C II}} / W_{\text{C IV}}$  with impact parameter was observed in 238 close galaxy pairs from the VIMOS Ultra Deep Survey \citep{H.mendez22}. The evolution of various ions also exhibits notable differences. According to \cite{dey2015}, an analysis of 31,727 absorption systems in JHU-SDSS revealed that the absorption of Fe\,\textsc{ii} and Mg\,\textsc{ii} follows a relation \( W_{\text{Fe II}} / W_{\text{Mg II}} \propto (-0.045 \pm 0.005)\thinspace z \). \citet{lan17} reported that the column density of C \,\textsc{i} decreases substantially at low redshift, with a more significant decline than observed in other ions (e.g., H\,\textsc{i}, C\,\textsc{ii}, Fe\,\textsc{ii}, Ni\,\textsc{ii}, Al\,\textsc{iii}, etc.).

%The formation, transport, and existence mechanisms of different elements vary significantly, leading to distinct distribution states, ionization conditions, and evolutionary paths in the CGM \citep{tumlinson17,peroux2024}. 
Despite extensive research, substantial uncertainties persist regarding the complex processes governing the CGM and the gas beyond dark matter halos. This paper builds upon the work of \cite{chen25}, with the goal of comprehensively deconstructing galaxy-associated gas in terms of its distribution and evolution. This undertaking is particularly challenging, as the formation, transport, and survival mechanisms of different ions can vary significantly, leading to  distinct distribution states, ionization conditions, and evolutionary paths \citep{tumlinson17,peroux2024}. We utilize  the Early Data Release (EDR) of the  Dark Energy Spectroscopic Instrument (DESI) to examine the equivalent width (\( W \)) of 33 absorption lines from 10 elements, analyzing spectra with foregrounds of QSOs across a redshift range of $0.3 < z <3.5$, and a impact parameter ($D$) ranging from 0.02 to 6.5 Mpc. %DESI aims at mapping the nature of dark energy with spectroscopic measurements of 40 million galaxies and quasars. Benefiting from its 5,000 optical fibers, DESI has unique advantages in large-scale observation capabilities \citep{levi19,abareshi22}. %Furthermore, the upcoming vast spectroscopic data from DESI Y1 (March 2025)  would provide  an unprecedented opportunity to probe the gas associated with galaxies.

This paper is structured as follows. Section \ref{sec:analysis} provides an overview of the DESI survey data, sample selection, and spectral stacking methodology. In Section \ref{sec:result}, we present key findings: the redshift-dependent evolutions of absorption line equivalent widths, their radial dependence on impact parameter, and the impact of ion type (Fe\,\textsc{ii} and Mg\,\textsc{ii}) on the distribution and evolution. Finally, Section \ref{sec:summary} synthesizes implications for gas evolution around galaxies and outlines future research plans. We adopt a flat $\Lambda$CDM cosmology with $h=0.677$,  $\Omega_{\rm M}=0.309$ and $\Omega_{\rm Lambda}=0.691$ using the \href{https://docs.astropy.org/en/latest/api/astropy.cosmology.realizations.Planck15.html}{\tt Planck15} package \citep{2016A&A...594A..13P}.

\section{DATA analysis} \label{sec:analysis}

\begin{figure*}
    \centering
    \includegraphics[width=2.1\columnwidth]{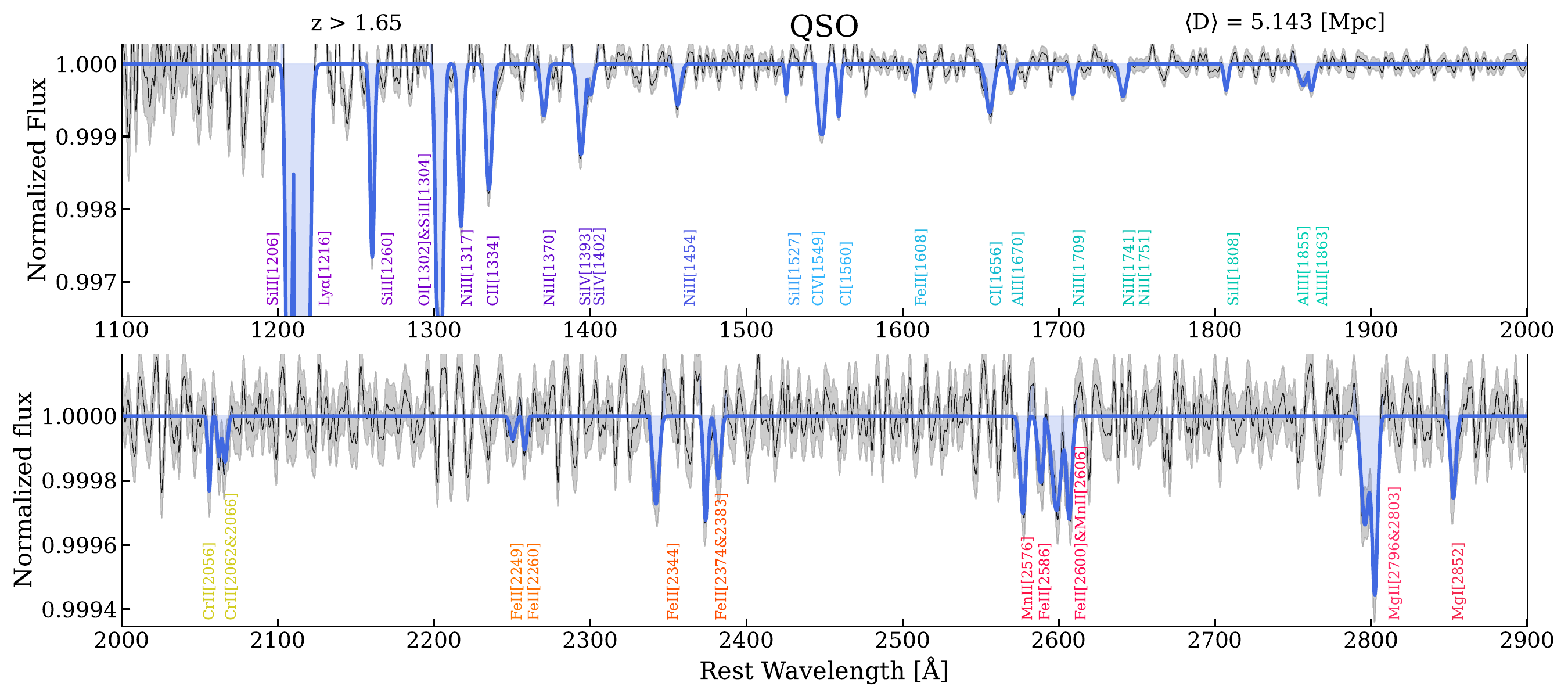}
    \caption{A normalized stacked spectrum of the QSOs. The black curve represents the co-added spectrum, while the blue curve depicts the Gaussian fit to the absorption lines. The shaded gray areas indicate the 1-$\sigma$ errors, and the shaded blue regions correspond to the equivalent widths  of the absorption lines. Text labels indicate the names of the corresponding absorption lines, and colors matching their rest wavelengths, as shown in the colorbar in Figure \ref{fig:metalev}. The plot also displays the redshift range (\(z > 1.65\))  and the mean impact parameter (\(\langle D \rangle= 5.1 \, \text{Mpc}\)) of the stacked spectrum. }
    \label{fig:spec}
\end{figure*}

\subsection{DESI EDR} \label{subsec:DESI}
The DESI \citep{aghamousa16a,aghamousa16b} is the largest multi-object spectrograph ever constructed, designed to conduct a large-scale spectroscopic survey to probe the nature of dark energy \citep{Dey19,abareshi22}. Installed on the 4-meter Mayall Telescope at Kitt Peak National Observatory,  DESI enables a survey of unprecedented scale, depth, and efficiency \citep{levi19}, mapping tens of millions of galaxies and quasars to study cosmic expansion. 

The Early Data Release\footnote{\url{https://data.desi.lbl.gov/doc/releases/edr/}} includes the spectra for 1.8 million unique targets observed during its Survey Validation phase (December 2020 -- June 2021; \citealt{adame23}). This phase tested DESI's instrument performance, refined target selection algorithms, and optimized observing procedures \citep{adame23,myers23}. The initial validation (SV1) used broad selection criteria \citep[e.g.,][]{guy2023} to calibrate DESI's spectroscopic pipelines and establish signal-to-noise requirements. Subsequent phases (SV2) and a ``One-Percent Survey" (SV3) demonstrated DESI's operational readiness by achieving high fiber assignment completeness and observational efficiency over a limited sky area. 

%\subsection{Datasets} \label{subsec:dataset}
Galaxies in DESI EDR can be categorized into three classes \citep{myers23}: Emission Line Galaxies \citep{raichoor20,raichoor23}, Luminous Red Galaxies  \citep{zhou20,zhou23}, and Bright Galaxy Survey (BGS) \citep{ruiz20,hahn23}. We utilize the {\tt{SPECTYPE}} from the {\tt zcatalog}\footnote{\url{https://data.desi.lbl.gov/public/edr/spectro/redux/fuji/zcatalog/zall-tilecumulative-fuji.fits}} of the DESI project to distinguish between QSOs \citep{yeche20,chaussidon23} and galaxies. The redshift estimation algorithm for galaxies in this catalog is known as {\tt redrock}\footnote{\url{https://github.com/desihub/redrock}}. Here, we apply a selection of \texttt{ZWARN} $= 0$ or $4$ to ensure the reliability of the redshift measurements in DESI. The stellar masses of galaxies are obtained from the value-added catalog \footnote{\url{https://data.desi.lbl.gov/public/edr/vac/edr/stellar-mass-emline/v1.0/edr_galaxy_stellarmass_lineinfo_v1.0.fits}} \citep{HuZou-19}, which estimates these masses using the CIGALE SED fitting code \citep{boquien19}. 

\subsection{Spectral stacking method} \label{subsec:method}

Following the work of \cite{chen25}, we conducted a positional cross-matching between QSOs and galaxies in the EDR, resulting in three types of pairs: QSO-QSO, QSO-ELG, and QSO-LRG pairs. In this paper, we focus primarily on the QSO-QSO pairs as QSOs cover a broader range of redshifts compared to galaxies, making them more suitable for studying cosmic metal enrichment. We stacked the background QSO spectra from these pairs at the rest-frame wavelengths of the foreground QSOs (or galaxies in Appendix \ref{sec:appenA}). The stacking method employed in this study closely parallels that of \cite{chen25}. Below, We provide a brief description of this method.
%At this time, we applied two iterations of a median filter to better mitigate fluctuations of varying degrees in the normalized spectra. The first iteration used a window size of 141 pixels (112.6 \AA), while the second iteration used a window size of 71 pixels (56.8 \AA). This stacking method is based on the work of \cite{zhu13, bordoloi11, zhu13ca, bordoloi14}; and \cite{lan18}, utilizing normalized spectra obtained from QSO spectra processed with the  \href{https://github.com/legolason/PyQSOFit}{code} from \cite{guo18}. 

We first performed continuum fitting of the background QSO spectra using the {\texttt{PyQSOFit}} software \footnote{\url{https://github.com/legolason/PyQSOFit}} developed by \cite{guo18}. The QSO spectra were then normalized by their continuum. %($\text{Flux}_{\text{norm}} = \text{Flux}/{\text{Flux}_{\text{conti}}}$)
Subsequently, we applied two iterations of a median filter to mitigate fluctuations in the normalized spectra. The first iteration employed a window size of 141 pixels ($\sim112.6$\,\AA), while the second iteration used a window size of 71 pixels ($\sim56.8$\,\AA). In both iterations, regions spanning 5 pixels ($\sim4$\,\AA) centered on the absorption line cores were masked to avoid contamination from the absorption features. 

The median-filtered spectra were then co-added onto a wavelength grid with an interval of 0.2 \AA, where the median value was adopted as the flux of the stacked spectrum. We employed the bootstrap resampling method to estimate the uncertainty of the normalized flux at each wavelength pixel, using 100 realizations. We compared the errors from 100 and 1000 bootstrap realizations and confirmed that the 100 realizations are sufficiently robustness for this work. Finally, we performed Gaussian fitting of the absorption features using the \texttt{emcee} \citep{foreman13} and integrated the flux within the absorption troughs to derive the equivalent width. This stacking approach is widely adopted as an unbiased method due to its flexibility and reliability \citep[e.g.,][]{bordoloi11,zhu13, zhu13ca,zhu14, bordoloi14}. Due to the broad wavelength range covered in this work, it is unavoidable that some absorption lines fall within the Ly$\alpha$ forest region of the background QSO. The potential impact of this mixed stacking is discussed in detail in Appendix~\ref{sec:appenC}. As shown, absorption line measurements are nearly unaffected by this effect. 

We present a stacked spectrum with a foreground QSO (\(z > 1.65\), \(\langle D \rangle = 5.1 \, \text{Mpc}\)) in Figure \ref{fig:spec}. The spectrum is divided into two segments (top and bottom panels) within the range of 1100 $< \lambda <$2900 \AA\, and an appropriate y-axis range is selected to simultaneously display absorption lines with varying intensities. As shown, after stacking hundreds of thousands of spectra, the fluctuations and errors in the stacked spectrum are minimized, enabling the detection of very faint absorption troughs. The complete spectral data are available in our publicly released \dataset[dataset]{https://doi.org/10.5281/zenodo.15617674}~\footnote{\url{https://doi.org/10.5281/zenodo.15617674}}.

\section{RESULTS} \label{sec:result}

We present the \( W \) (in m\AA) of all detected absorption lines using QSOs as foreground samples in Table \ref{tab1}. %and Table \ref{tab2}). 
The data are  organized into four redshift bins: $z < 0.75$, $0.75 < z < 1.2$, $1.2 < z < 1.65$, and $z > 1.65$, along with three impact parameter bins: $0.02\thinspace \mathrm{Mpc}< D  < 1.2\thinspace \mathrm{Mpc}$, $1.2\thinspace \mathrm{Mpc}< D  < 3.5\thinspace \mathrm{Mpc}$, and $3.5\thinspace \mathrm{Mpc}< D  < 6.5\thinspace \mathrm{Mpc}$. The average impact parameters  for three bins are $\langle D \rangle=$ 0.8 Mpc, 2.5 Mpc, and 5.1 Mpc, respectively. A blank entry indicates that the absorption line wavelength exceeds the limits of the co-added spectrum, while a dash denotes cases where the $S/N$ is too low to fit the absorption line and measure the \(W\). 

Additionally, an asterisk (\thinspace*\thinspace) in Table~\ref{tab1} indicates either measurements with $S/N < 3$ or absorption features identified as narrow, jitter-induced depressions misinterpreted by Gaussian fitting. Gaussian fitting may occasionally capture narrow spurious absorption features caused by spectral fluctuations (e.g., Fe\,\textsc{ii}\,$\lambda$\,2249 and Fe\,\textsc{ii}\,$\lambda$\,2260 in Figure~\ref{fig:spec}). To mitigate this, we impose a criterion that the equivalent width of a line must exceed three times the square root of the summed squared error within a 5\,\AA\ window centered on the line. This window is broad enough to suppress spurious features from narrow fluctuations while retaining genuine absorption lines. Nonetheless, final visual inspection remains essential to ensure the reliability of the fits. Measurements of absorption lines that lie too close to the blue end of the stacked spectrum, where spectral coverage becomes insufficient, are also struck through with a horizontal line and excluded from the analysis.  

Due to space limitations, errors for \( W \) are not included here; however, a complete table with errors is available in our dataset. In this study, we select QSOs as foreground pairs for the main sample, as they provide a broad redshift distribution of foreground objects, facilitating a more comprehensive examination of evolutionary effects. The equivalent width results for ELGs and LRGs are included in Appendix \ref{sec:appenA}, and the corresponding data, including the stacked spectra and absorption measurements, can be accessed on the same dataset. The impact parameter bins, ranging from 0.02 to 6.5 Mpc, encompass the CGM  and the gas that connects to the large-scale structure of the universe, making them suitable for conducting a thorough survey of the gas associated with QSOs.

We measured \(W\)  of a total of 33 absorption lines around QSOs (and galaxies). These absorption lines originated from ten elements (Si, O, C, Al, Mn, Ly$\alpha$, Ni, Fe, Cr, and Mg), tracing gas from cold (e.g., Mg\,\textsc{i}, C\,\textsc{i}; $\sim10^3 \, \text{K}$) to hot (e.g., C\,\textsc{iv}, Si\,\textsc{iv}; $\sim10^5 \, \text{K}$) and from low (e.g., Mg\,\textsc{ii}, Fe\,\textsc{ii}) to high (e.g., C\,\textsc{iv}, Si\,\textsc{iv}) ionization states. By comparing different redshift and impact parameter bins, we can gain insights into the distribution and evolution of gas around galaxies. Additionally, various ions serve as tracers of different gas components, enhancing our understanding of the complex multiphase nature of these gases. Thus, the data presented in Table \ref{tab1} contains a wealth of information regarding the gas around quasars and its connection to the large-scale structure, making it highly valuable.

%Table \ref{tab2} is similar to Table \ref{tab1}, but it uses ELGs and LRGs as foreground samples. The four redshift bins are: ELG \(z < 1.1\), ELG \(z > 1.1\), LRG \(z < 0.7\), and LRG \(z > 0.7\). Due to the size limitations of the tables, the error for \( W \) are not included here. So, We place the complete table in this \href{https://cloud.ecwang.top:8443}{website}.

%\clearpage
\begin{longtable*}{p{2.7cm}|p{1cm}p{1cm}p{1cm}p{1cm}p{1cm}p{1cm}p{1cm}p{1cm}p{1cm}p{1cm}p{1cm}p{1cm}}
% 表格内容
\caption*{$W$ in  (m\AA) for different ions around QSOs at different redshifts} \label{tab1}\\

\hline
\hline
\multicolumn{1}{c}{\textbf{Redshift}} & \multicolumn{3}{c}{\textbf{$z < 0.75$}} & \multicolumn{3}{c}{\textbf{$0.75 < z < 1.2$}} & \multicolumn{3}{c}{\textbf{$1.2 < z < 1.65$}} & \multicolumn{3}{c}{\textbf{$z > 1.65$}}  \\

\multicolumn{1}{c}{\textbf{$<$D$>$ [Mpc]}} & 0.8 & 2.5 & 5.1 & 0.8 & 2.5 & 5.1  & 0.8 & 2.5 & 5.1 & 0.8 & 2.5 & 5.1 \\
\hline

\endhead

Si\,\textsc{ii}[1206] &  &  &  &  &  &  &  &  &  & 42.2 & 29.8 & 44.0 \\
Ly$\alpha$[1216] &  &  &  &  &  &  &  &  &  & 331 & 196 & 167 \\
Si\,\textsc{ii}[1260] &  &  &  &  &  &  &  &  &  & 30.1 & 11.5 & 8.79 \\
O\,\textsc{i}[1302]\&Si\,\textsc{ii}[1304] &  &  &  &  &  &  &  &  &  & 26.0 & 22.0 & 22.3 \\
Ni\,\textsc{ii}[1317] &  &  &  &  &  &  &  &  &  & 25.6 & 11.4 & 8.82 \\
C\,\textsc{ii}[1334] &  &  &  &  &  &  &  &  &  & 13.4 & 15.4 & 8.85 \\
Ni\,\textsc{ii}[1370] &  &  &  &  &  &  & – & – & – & – & 5.41 & 3.23 \\
Si\,\textsc{iv}[1393] &  &  &  &  &  &  & $\sout{128}$ & $\sout{135}$ & $\sout{128}$ & 11.4 & 9.47 & 7.21 \\
Si\,\textsc{iv}[1402] &  &  &  &  &  &  & $\sout{127}$ & $\sout{63.5}$ & $\sout{80.8}$ & 14.8 & 5.51 & 3.86 \\
Ni\,\textsc{ii}[1454] &  &  &  &  &  &  & $\sout{39.7}$ & $\sout{33.5}$ & $\sout{28.3}$ & 5.53 & 2.00 & 2.72 \\
Si\,\textsc{ii}[1527] &  &  &  &  &  &  & $\sout{4.09}$* & $\sout{17.9}$ & $\sout{10.4}$ & 1.73* & 2.94 & 0.83 \\
C\,\textsc{iv}[1549] &  &  &  &  &  &  & 39.4 & 23.0 & 15.2 & 21.6 & 8.48 & 5.50 \\
C\,\textsc{i}[1560] &  &  &  &  &  &  & 5.99 & 7.94 & 9.48 & – & – & 1.82 \\
Fe\,\textsc{ii}[1608] &  &  &  &  &  &  & 15.5 & 8.97 & 8.31 & 4.15 & 0.72* & 0.85* \\
C\,\textsc{i}[1656] &  &  &  & $\sout{286}$ & $\sout{349}$ & $\sout{308}$ & – & 3.44 & 2.69 & 5.87 & 4.76 & 3.69 \\
Al\,\textsc{ii}[1670] &  &  &  & $\sout{117}$ & $\sout{161}$ & $\sout{132}$ & 11.4 & 6.29 & 3.15 & 6.79 & 2.28 & 1.23 \\
Ni\,\textsc{ii}[1709] &  &  &  & $\sout{86.0}$ & $\sout{45.4}$ & $\sout{53.6}$ & 5.69 & 3.41 & 2.12 & 2.25* & 2.50 & 1.36 \\
Ni\,\textsc{ii}[1741] &  &  &  & $\sout{58.7}$ & $\sout{49.4}$ & $\sout{43.6}$ & 0.01* & 0.73* & 3.27 & 3.19* & 1.73* & 2.46 \\
Ni\,\textsc{ii}[1751] &  &  &  & $\sout{46.0}$ & $\sout{34.1}$ & $\sout{28.5}$ & 2.07 & 2.83 & 1.15 & – & 1.03* & – \\
Si\,\textsc{ii}[1808] &  &  &  & 27.6 & 22.7 & 19.3 & 1.64* & 1.57 & 1.07 & – & 3.14 & 1.08 \\
Al\,\textsc{iii}[1855] &  &  &  & 10.5* & 16.0 & 13.6 & 9.63 & 1.57 & 3.31 & 6.42 & 2.03 & 2.59 \\
Al\,\textsc{iii}[1863] &  &  &  &5.09* & 8.78 & 10.5 & 5.01 & 1.48 & 1.52 & 0.71* & 1.59 & 2.07 \\
Cr\,\textsc{ii}[2056] &  &  &  & 0.14* & 2.77 & 1.92 & 0.04* & 0.63* & 0.41* & 2.18* & 1.30 & 0.48 \\
Cr\,\textsc{ii}[2062\&2066] & – & – & – & 4.63* & 2.50 & 1.73 & 9.31 & 0.92* & 1.20* & 0.01* & 1.4* & 1.01 \\
Fe\,\textsc{ii}[2249] & 29.2 & 21.3 & 25.2 & – & 1.48 & 1.97 & – & 0.64* & 1.89 & – & 1.04* & 0.33* \\
Fe\,\textsc{ii}[2260] & 39.6 & 15.4 & 22.4 & – & 2.16 & 0.32* & 6.97 & 1.84 & 1.41 & 3.36 & 0.26* & 0.30* \\
Fe\,\textsc{ii}[2344] & 39.9 & 7.23 & 17.0 & 5.24 & 1.69 & 1.23 & 5.58 & 1.84 & 1.91 & 1.73* & 1.94 & 1.18 \\
Fe\,\textsc{ii}[2374\&2383] & 51.2 & 40.3 & 31.7 & 4.64 & 2.60 & 4.08 & 3.12* & 3.30 & 1.96 & 5.25 & 1.55 & 1.89 \\
Mn\,\textsc{ii}[2576] & 29.2 & 16.8 & 9.73 & 1.38* & 3.06 & 1.33 & 5.08 & 0.30* & – & 0.90* & 0.42* & 1.36 \\
Fe\,\textsc{ii}[2586] & 21.3 & 21.1 & 15.9 & 3.45 & 2.35 & 1.46 & – & 2.60 & 0.83 & – & 0.77* & 0.93 \\
{\footnotesize Fe\,\textsc{ii}[2600]\&Mn\,\textsc{ii}[2606]} & 23.8 & 30.5 & 37.5 & 4.20* & 2.64 & 4.15 & 3.40* & 3.17 & 3.31 & 12.9 & 2.75 & 3.37 \\
Mg\,\textsc{ii}[2796\&2803] & 26.6 & 7.47 & 6.83 & 13.8 & 3.74 & 5.04 & 12.5 & 3.88 & 2.32 & 20.6 & 5.38 & 4.16 \\
Mg\,\textsc{i}[2852] & 1.70* & 1.94* & 5.07 & 2.73* & 3.45 & 1.77 & 2.66* & 1.29 & 0.90 & 0.43* & 1.24 & 0.98 \\
\hline
\hline

\end{longtable*}

%\clearpage
\subsection{Evolution of $W$ with redshift and enrichment rates} \label{subsec:evo}

\begin{figure*}
    \centering
    \includegraphics[width=2\columnwidth]{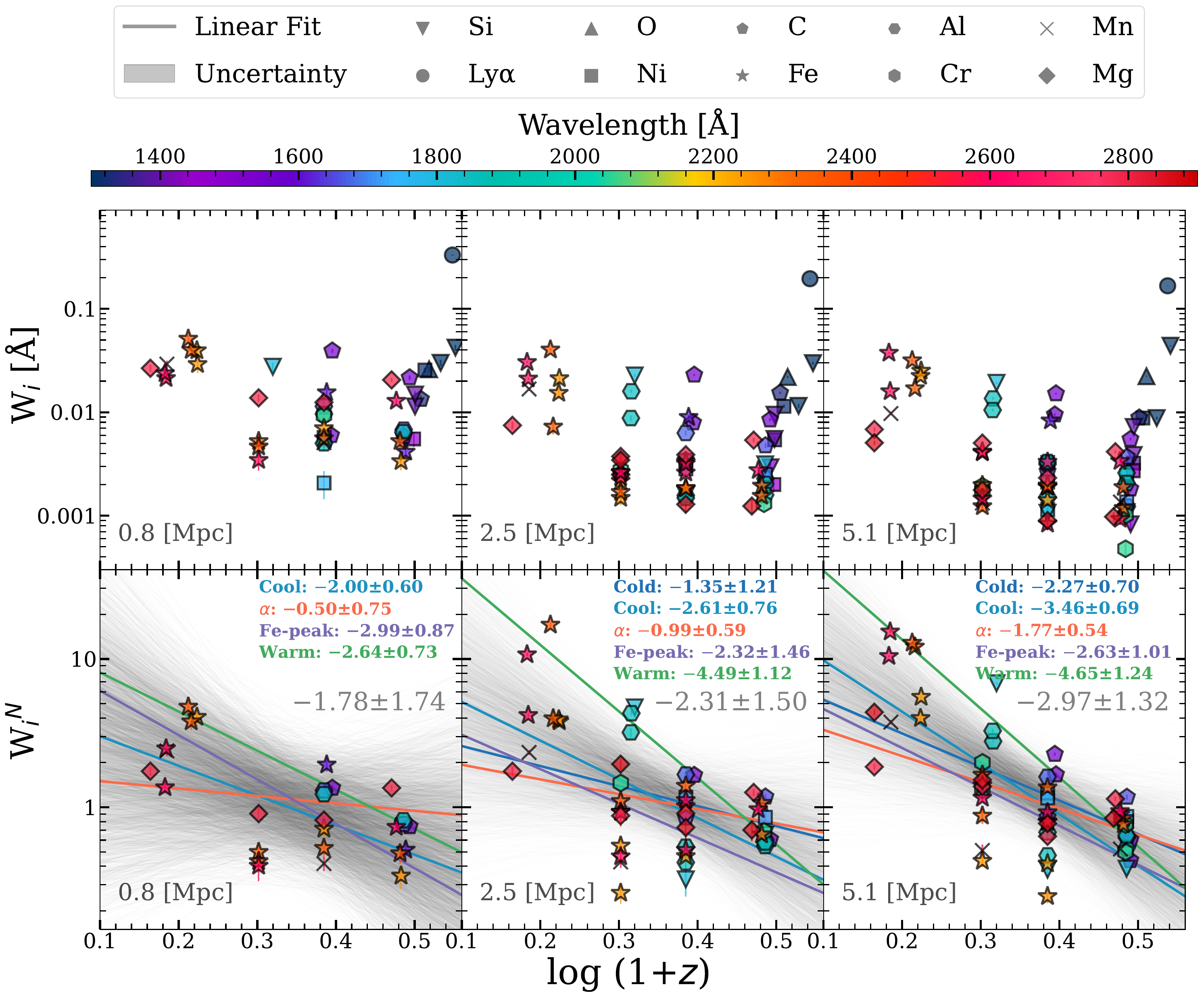}
    \caption{Top panels:  $W_i$ plotted against $\log(1+z)$ for three radial bins ($\langle D \rangle=$ 0.8 Mpc, 2.5 Mpc, and 5.1 Mpc). The marker shapes denote different elements (e.g., \texttt{Ly$\alpha$}\scalebox{1.3}{$\bullet$}, 
    \texttt{Si}$\blacktriangledown$, \texttt{O}$\blacktriangle$, \texttt{Ni}\scalebox{0.75}{$\blacksquare$}, etc.), while the colors correspond to their rest wavelengths. Error bars represent 1-$\sigma$ uncertainties in the measurements. Bottom panels: normalized equivalent widths \( W^N = W / 10^{ \langle \log W \rangle}\) as a function of $\log(1+z)$. The gray lines in the lower panels represent a linear fit and its associated uncertainty range. And the colored lines represent fits to specific types of absorption lines based on their tracing gas or the origin mechanisms of the elements, with corresponding slopes and uncertainties indicated in the text of the same color. The  observed \( W_i\) decreases with increasing redshift, which indicates the enrichment of the gas associated with QSOs. }
    \label{fig:metalev}
\end{figure*}

We plot the data from Table~\ref{tab1} that are not marked with an asterisk and are sufficiently removed from the blue end of the stacked spectrum in Figure~\ref{fig:metalev}. This selection criterion is consistently applied throughout all figures in this paper.
%To ensure the robustness of the plotted points, we require that each absorption line has at least two equivalent width measurements within four redshift bins. 
The upper panels display the equivalent widths (\( W_i \)) using different marker shapes to represent each element, with colors indicating their respective rest wavelengths. This marking scheme is consistent across all figures in this work, except for Figure \ref{fig:femg}, which focuses on the ratio of two lines. The lower panels of Figure \ref{fig:metalev} present the normalized equivalent widths (\( W_i^N = W_i / 10^{ \langle \log W_i \rangle}\)) as a function of \( \log(1+z) \).

A consistent negative correlation is observed across all ions, where \( W_i \) systematically decreases as redshift increases. The lower panels illustrate this trend more distinctly. The normalization of \( W_i^N\) accounts for differences in the initial \( W_i \) values, facilitating clearer comparisons of trends across various radial bins and redshifts. The observed decrease in \( W_i \) with redshift is evident in all radial bins with slopes of $\alpha = -1.78^{\pm 1.74}$ (0.8 Mpc), $-2.31^{\pm 1.50}$  (2.5 Mpc), and $-2.97^{\pm 1.32}$ (5.1 Mpc). 

We also fit the evolution separately for ions tracing cold gas (e.g., C\,\textsc{i}, Mg\,\textsc{i}), cool gas (e.g., Fe\,\textsc{ii}, Mg\,\textsc{ii}), and warm gas (e.g., C\,\textsc{iv}, S\,\textsc{iv}), as well as for $\alpha$-process elements (e.g., Mg, O, Si) and Fe-peak elements (e.g., Fe, Ni, Mn). Variations in the nucleosynthetic origins of these elements may lead to distinct evolutionary trends over cosmic time. In general, $\alpha$-elements show slightly weaker evolution at $z < 2$ compared to Fe-peak elements, while absorption lines tracing warmer gas exhibit somewhat stronger evolution. Nevertheless, given the relatively large uncertainties in the fitted slopes, these comparisons should be interpreted with caution.

% Galactic feedback is crucial in shaping the multiphase structure of the CGM at distances less than $0.5\, r_{200c}$, whereas regions beyond this are more influenced by cosmological effects, such as non-spherical accretion of intergalactic gas \citep{fielding20,li21}. This trend suggests that over cosmic time, metals produced and released by both star formation  processes and Active Galactic Nuclei (AGN) feedback contribute to the overall enrichment of metals in the universe \citep[e.g.,][]{moll07,germain09,huang23,mart24}. Initially, this enrichment occurs in the gas closest to QSOs (or galaxies), and as the inner regions approach a state of near-complete metal enrichment, these metals gradually propagate outward into more distant regions.  Consequently, in the later universe, we are more likely to observe enrichment processes taking place in the outer regions. This dynamic also accounts for the faster enrichment at larger $D$, as there is a delay in the propagation of metal enrichment from small to large scales associated with galaxies. 
%It also explains why the evolution of  at larger  $D$ is faster, as they interact directly with the cosmic web and the intergalactic Medium (IGM). 
% XXXX I did not get the logic of this explanation. 

\begin{figure*}
    \centering
    \includegraphics[width=2\columnwidth]{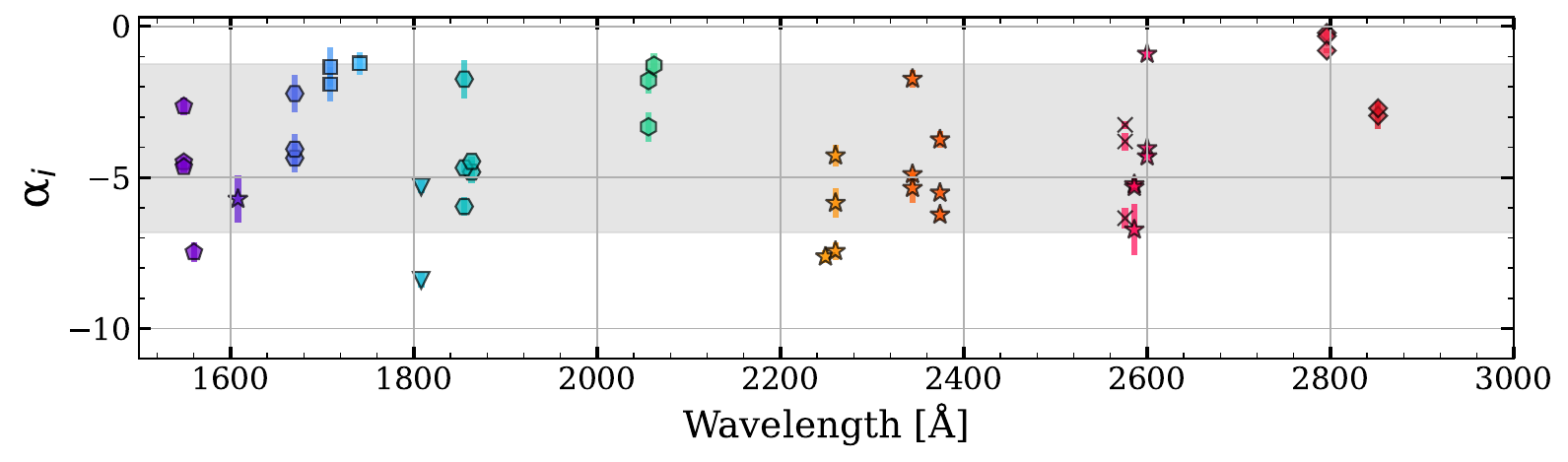}
    \caption{Fitted slopes $\alpha$ for each ion in the $\log(W)$ vs. $\log(1+z)$ plot (depicted in Figure \ref{fig:metalev}) shown against a rest wavelength range from 1500 \(\mathrm{\AA}\) to 3000 \(\mathrm{\AA}\). The colors and marker shapes adhere to the same convention as in Figure \ref{fig:metalev}. The center of the shaded region represents the average value of $\alpha$ for data points and the width indicates two times the standard deviation.}
    \label{fig:alpha}
\end{figure*}

In Figure \ref{fig:alpha}, we present the fitted slopes $\alpha$ for each absorption line in the \( \log W \) vs. \( \log(1 + z) \) plot (as shown in Figure \ref{fig:metalev}). The evolution of different elements and ions shows significant variability. For example, among all the absorption lines, Mg\,\textsc{ii} shows the weakest evolution, indicating little to no change over time, which aligns with trends noted by \cite{lan20}, \cite{anand21} and \cite{chen25}. The overall structure of Mg\,\textsc{ii}-bearing halos was established early in the galaxy assembly process and have evolved relatively little since then \citep{matejek12}.

\subsection{$W$ changing with impact parameter}\label{subsec:impact parameter}

\begin{figure*}
    \centering
    \includegraphics[width=2\columnwidth]{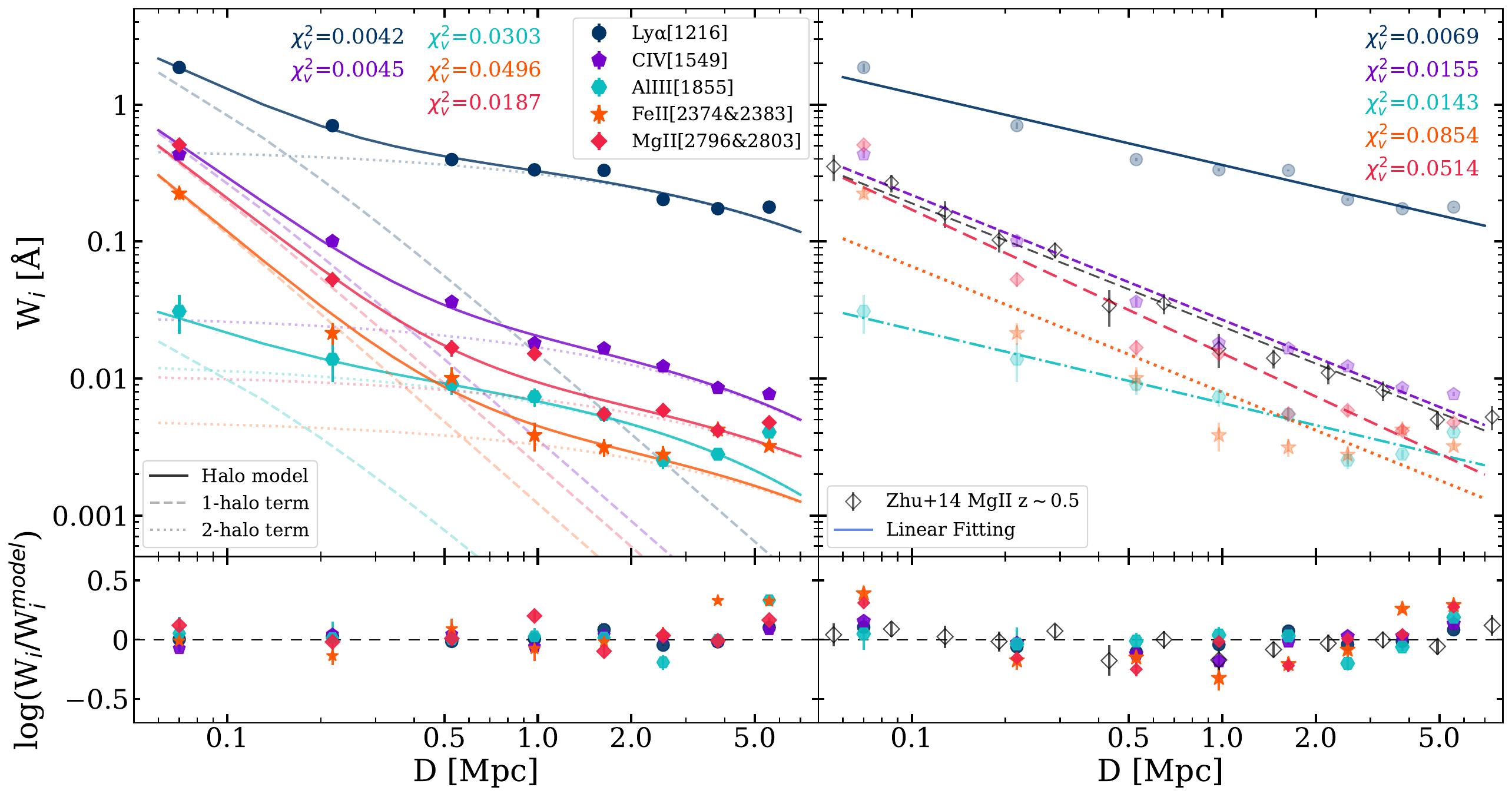}
    \caption{Top left: $W_i$  as a function of the impact parameter $D$ for various ions. Different colors and marker shapes correspond to different absorption lines, following the same convention as in Figure \ref{fig:metalev}.  The data points are fitted using the Halo model (solid line) with 1-halo term (dashed line) and 2-halo term (dotted line). Top right: Linear fitting of the observed data; hollow black diamond markers indicate the Mg\,\textsc{ii} absorption for LRGs as foreground galaxies, as detailed by \cite{zhu14}. The reduced chi-squared values ($\chi^2_\nu$) for both the halo model and the linear fit are also shown in the figure to compare the goodness of fit between the two models. Bottom left and bottom right: logarithmic ratio \( \log(W_i/W_{i}^{\text{model}}) \) of the observed $W_i$ to the predicted values from the model ($W_{i}^{\text{model}}$). Error bars indicate the uncertainty of $W_i$.}
    \label{fig:D}
\end{figure*}

\begin{figure*}
    \centering
    \includegraphics[width=2.1\columnwidth]{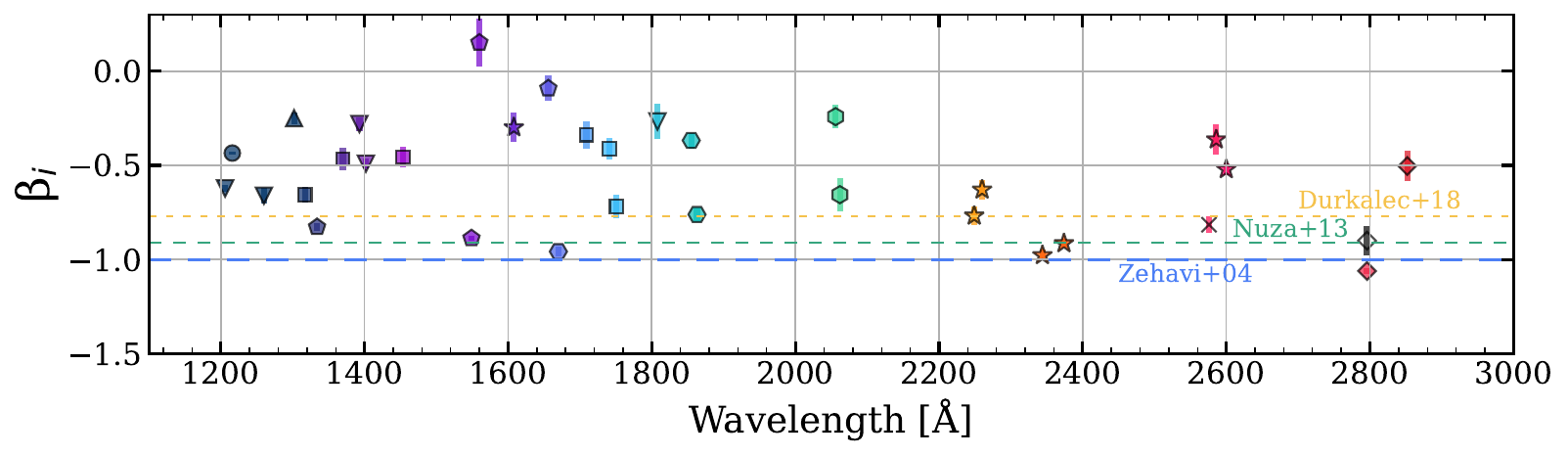}
    \caption{Fitted slopes $\beta$ for each ion in the $\log(W)$ vs.\ $\log(D)$ plot (Figure \ref{fig:D}) shown as a function of rest wavelength (1200\,\AA\ to 3000\,\AA). The colors and marker shapes follow the same convention as in Figure \ref{fig:metalev}. Hollow black diamond markers represent the Mg\,\textsc{ii} slope from \cite{zhu14}. Projected two-point correlation function slopes of galaxies within the range of 1–5 Mpc are indicated: \cite{zehavi04} for \( 0.02 < z < 0.16 \) (blue), \cite{nuza13} for \( 0.4 < z < 0.7 \) (green), and \cite{durkalec18} for \( 2 < z < 3.5 \) (orange).}
    \label{fig:beta}
\end{figure*}

In Figure \ref{fig:D}, we investigate the radial distribution of  \( W_i \) for various absorption lines, including Ly\(\alpha\)$\thinspace \lambda 1216$, C\,\textsc{iv}$\thinspace \lambda 1549$, Al\,\textsc{iii}$\thinspace \lambda 1855$, Fe\,\textsc{ii}$\thinspace \lambda\lambda 2374, 2383$, and Mg\,\textsc{ii}$\thinspace \lambda\lambda 2796, 2803$. To enhance the radial sampling resolution, we combine QSOs from four redshift bins and categorize them into eight impact parameter bins with boundaries at [0.02, 0.1, 0.3, 0.7, 1.2, 2.0, 3.0, 4.5, 6.5] Mpc.

Building on the work of \cite{zhu14} and \cite{wu2024}, we fit the \( W_i \) of these gas halos, observing a distribution that exhibits a two-halo behavior (left panel in Figure \ref{fig:D}).  The two-halo term describes the distribution of gas in large-scale environments beyond individual galaxy (or QSO) halos, influencing galaxy formation. This term accounts for the correlation between different halos and is governed by the halo-halo correlation function \citep[see][for details]{zhu14}. Specifically, the two-halo term considers the large-scale clustering of gas, typically modeled using linear bias theory, and incorporates the halo mass function and the cosmic power spectrum.

We also perform the linear fittings to the $W_i$ profiles of these absorption lines as shown in the right panel of Figure \ref{fig:D}. The results indicates that the absorption intensity decreases with $D$, which aligns with previous findings \citep[e.g.,][]{zhu13ca,lan18,H.mendez22,wu2024}. Mg\,\textsc{ii} absorption strength here is slightly lower than \cite{zhu14} for LRGs at $z\sim0.5$. This discrepancy primarily arises from the higher redshift of the foreground QSOs ($z\sim1.5$) in our study.  Ly\(\alpha\) demonstrates a stronger absorption as predicted by \cite{shen13} and remains relatively stable on large scales. 
Even after accounting for increased model complexity, the halo model still provides a better overall fit (with a smaller $\chi^2_\nu$) than the linear model. In Figure \ref{fig:D}, the data points at 5 Mpc appear higher than both models. This may be due to the effect of QSO proximity zones \citep{Zheng15, Onorato25}, beyond which the ionization influence of the host QSO diminishes. This reduction in ionization could lead to an increase in the fraction of low-ionization species.

The fitted slopes \( \beta \) from the linear fittings for each ion are displayed in Figure \ref{fig:beta}.
The observed differences in $\beta$ could be attributed to the varying physical processes that affect the distribution of each ion in the gas surrounding galaxies. For example, high-ionization species such as C\,\textsc{iv} and Si\,\textsc{iv} are typically found in hotter, more ionized regions. In contrast, low-ionization species such as Mg\,\textsc{ii} and Fe\,\textsc{ii} are generally located in colder, and denser regions. High-ionization species tend to maintain a more extended distribution (with smaller $|\beta|$), as corroborated by both simulations \citep{shen13,sanchez19,li21,carr25} and observational studies \citep{tumlinson11,prochaska11,H.mendez22}. Conversely, low-ionization species Mg\,\textsc{ii} and Fe\,\textsc{ii} show stronger dependence on impact factor, with $\beta$ values of $-0.77$ (Fe\,\textsc{ii} \(\lambda 2249\)), $-0.63$ (Fe\,\textsc{ii} \(\lambda 2260\)), $-0.98$ (Fe\,\textsc{ii} \(\lambda 2344\)), $-0.91$ (Fe\,\textsc{ii} \(\lambda\lambda 2374, 2383\)), and $-1.06$ (Mg\,\textsc{ii} \(\lambda\lambda 2796, 2803\)). Additionally, in Figure \ref{fig:beta}, we present the slope of the projected two-point correlation function for galaxies within the 1–5 Mpc range, marked by different colors: \cite{zehavi04} for \( 0.02 < z < 0.16 \) (blue), \cite{nuza13} for \( 0.4 < z < 0.7 \) (green), and \cite{durkalec18} for \( 2 < z < 3.5 \) (orange).

The radial profile of the absorption line equivalent width \( W_i(D) \) could serve as a direct tracer of the spatial distribution of multi-phase gas associated with QSOs and galaxies. However, its continuous nature and the complex physical processes, such as feedback and ionization, distinguish it from the discrete clustering of galaxy-galaxy correlations \citep[e.g.,][]{davis1983,baugh1996,zehavi04,eisenstein05,Li06,nuza13,durkalec18}. The projected two-point correlation function \( w_p(D) \) for galaxies follows the relation \( w_p \propto D^\gamma \), where \( \gamma \) varies depending on the scale, redshift, and galaxy selection. At a scale of \( \sim 5 \, \text{Mpc} \) and redshift  \( 2 < z < 3.5 \) (from the VIMOS Ultra Deep Survey), \( \gamma \) is roughly $-0.8$ \citep{durkalec18}. This value is comparable to the \( \beta \) values of the absorption lines tracing cold gas, suggesting that cold gas may significantly influence environments conducive to galaxy formation. 

The continuous distribution of gas is shaped by a combination of large-scale cosmic forces and local interactions, such as stellar feedback and gas ionization. This creates a complex and evolving relationship between the gas distribution and the position of galaxies within their environments. In contrast, galaxy-galaxy correlations tend to present a more discrete and static clustering pattern. Therefore, it is important to exercise caution when making direct comparisons between these two phenomena. Despite these differences, we argue that the radial profiles of absorption lines can serve as valuable tracers of multi-phase gas environments. They complement the information provided by two-point correlation functions. Future large spectroscopic surveys, including the DESI survey, will significantly enhance the application of this method, allowing for a deeper understanding of the interplay between gas and galaxy formations.

\subsection{$W$ ratio of different elements}\label{subsec:Ionization}

As discussed previously, the evolution and distribution of different elements and their various ionization states (e.g. C\,\textsc{i} vs. C\,\textsc{ii}, \cite{lan17}; C\,\textsc{ii} vs. C\,\textsc{iv}, \cite{H.mendez22}), exhibit significant variation. We illustrate this using Mg\,\textsc{ii} and Fe\,\textsc{ii} as examples. These two ions trace gas with similar temperature and ionization states. The first ionization potentials for Mg and Fe are closely aligned, with values of 7.646 eV for Mg\,\textsc{i} → Mg\,\textsc{ii} and 7.902 eV for Fe\,\textsc{i} → Fe\,\textsc{ii}. The second ionization potentials (Mg\,\textsc{ii} → Mg\,\textsc{iii} and Fe\,\textsc{ii} → Fe\,\textsc{iii}) are 15.0357 eV and 16.1878 eV, respectively \citep{morton03,kramida13}.

\begin{figure*}
    \centering
    \includegraphics[width=2\columnwidth]{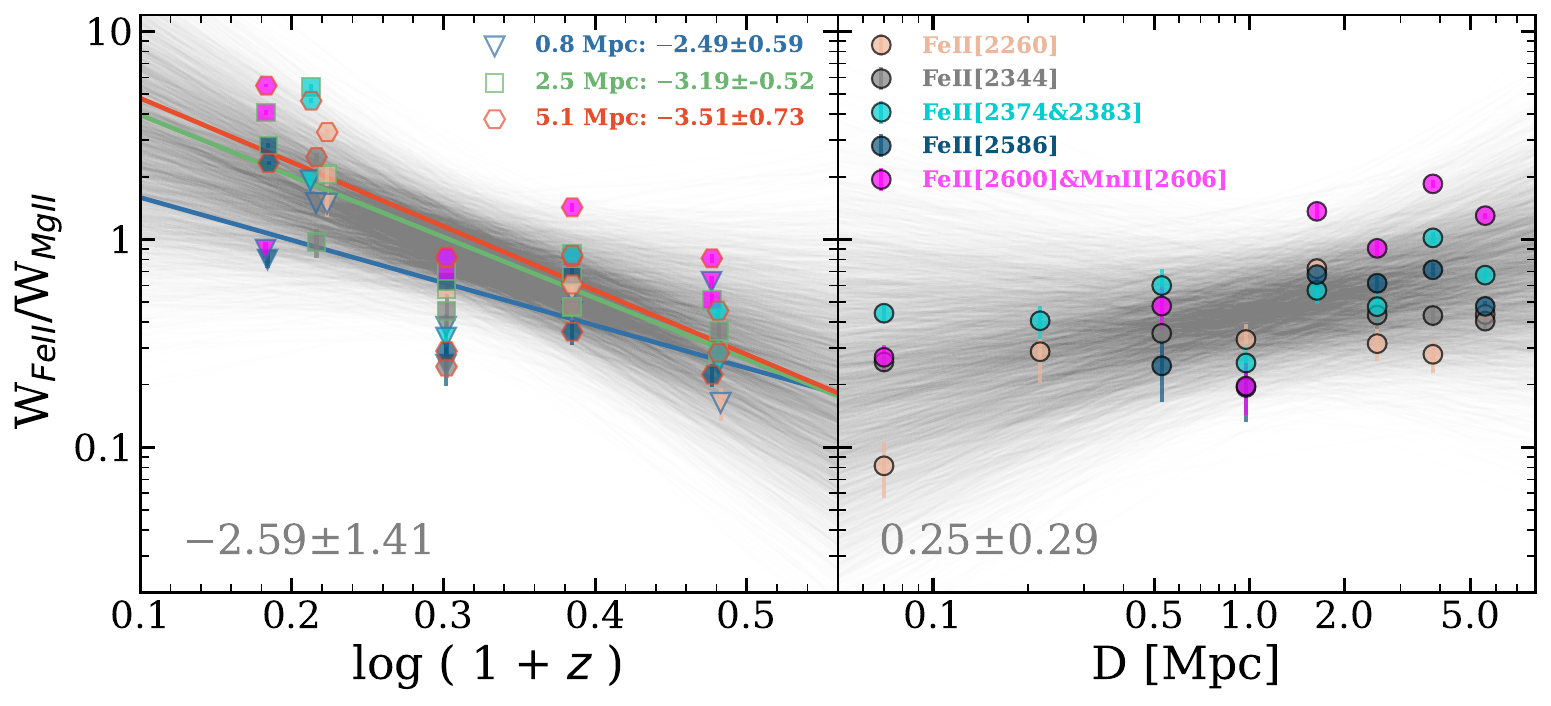}
    \caption{The $W_{\mathrm{FeII}} / W_{\mathrm{MgII}}$ ratio plotted against $\log(1+z)$ (left) and the  $D$ (right). Different colors represent various FeII lines, as indicated in the legend, with markers added in the left panel to show the corresponding impact parameter bins. The gray lines depict linear fits along with their uncertainties. We also fit data in different impact parameter bins separately, with slopes and uncertainties labeled in corresponding colors. The data clearly show that the $W_{\mathrm{FeII}} / W_{\mathrm{MgII}}$ ratio decreases with increasing redshift \(z\). 
}
    \label{fig:femg}
\end{figure*}

The ratio \( W_{\text{FeII}} / W_{\text{MgII}} \) is shown as a function of \( \log(1 + z) \) (left panel) and the impact parameter (right panel) in Figure \ref{fig:femg}. Different colors correspond to various Fe\,\textsc{ii} absorption lines indicated in the legend. The results indicate that as the redshift increases, the ratio \( W_{\text{FeII}} / W_{\text{MgII}} \) decreases following a trend proportional to $\propto (1+z)^{-2.59\pm 1.41}$, consistent with \cite{dey2015}. This trend is consistently observed within each individual impact parameter bin. In contrast, the variation of \( W_{\text{FeII}} / W_{\text{MgII}} \) with distance is not significant within the 1-$\sigma$ confidence level.

%Conversely, as \( D \) increases, the ratio shows a slight increase. This suggests that Mg\,\textsc{ii} is likely more concentrated in the inner regions and evolves more slowly with redshift. 

Mg, being an $\alpha$-element, is predominantly produced by core-collapse supernovae (CCSN). In contrast, iron (Fe) is primarily generated by Type Ia supernovae (SNIa), which originate from the explosion of accreting white dwarfs. The ratio of these elements should reflect the relative frequencies of CCSN and SNIa, as well as the time lag between their productions. In studying the stellar absorption lines, \cite{conroy-14} found significant $\alpha$ enhancements in massive elliptical galaxies formed in the early universe, attributed to the intense bursts of star formation that occurred at high redshift \citep{Li-25}. 

%At high redshift (early universe), CCSN contribute more, leading to Mg enrichment and a lower [Fe/Mg]. At low redshift (late universe), SNIa gradually contribute more Fe, causing the [Fe/Mg] to increase. This is consistent with the theoretical predictions of SNIa/CCSN rate evolution by \cite{maoz12} and \cite{madau14}.

In addition to the delay in element production, variations in the initial mass function (IMF) could also influence the [Fe/Mg]. Several studies suggest that the IMF at higher redshifts may be top-heavy. For exanple, a low \( {^{13}\text{CO}}/{\text{C}^{18}\text{O}} \) ratio has been observed in strongly lensed starbursts at \( z \sim \) 2-3  \citep{danielson13,yang2023}, which can be effectively modelled in galactic chemical evolution (GCE) models assuming a top-heavy IMF \citep{zhang2018}. This low ratio has been reported for high-redshift ($z\approx2.26$) main-sequence galaxies \citep{guo2024}, further supporting the idea of a top-heavy IMF.  A top-heavy IMF increases the proportion of massive stars, elevating the number of CCSN, enhancing Mg production, and consequently lowering the [Fe/Mg]. This effect is also reflected in the simulation by \cite{palla2020}.

% In contrast, Fe is predominantly produced by SNIa, where the progenitor stars are white dwarfs in binary systems. These white dwarfs undergo accretion or merger processes that typically take about 1 Gyr \citep{matteucci01,maoz10,horiuchi10}. This delay decouples Fe release from the star formation activities associated with dense gas clouds, allowing Fe to be transported to more distant regions following supernova explosion. This may explain the increase \( W_{\text{FeII}} / W_{\text{MgII}} \)  with  \( D \). Additionally, on small scales, Fe may be partially adsorb onto dust particles \citep{Konstantopoulou22}, reducing its observed abundance in ionized gas. As the impact parameter\( D \) increases, the dust density decreases, causing Fe to desorb from the dust and re-enter the gas phase, which may contribute to an increase in the [Fe/Mg] ratio.

\section{Summary} \label{sec:summary}

This study investigates the spatial distribution and redshift evolution of metal species in the gas associated with QSOs using the early data release from the DESI. We analyze 33 absorption lines from 10 elements within the redshift range $0.3 < z < 3.5$ and impact parameters $D < 6.5$ Mpc. These absorption lines provide insights into the diffuse presence of various elements in different states, offering valuable information on structure growth, cosmic metal enrichment, and thermodynamic environments.  The behavior of these gases is complex and closely related to various ions and absorption. Our main findings are as follows:

\textbf{Cosmic Evolution:}
The absorption strength of various elements and ions systematically decreases with increasing redshift, following the trend $W \propto (1+z)^{-4.0 \pm 2.7}$ (Figure \ref{fig:alpha}). This behavior indicates a progressive enrichment of metals over cosmic time. The enrichment rate is closely related to the production mechanisms and ionization states of the elements. Specifically, Fe-peak elements appear to exhibit somewhat stronger evolution than $\alpha$-elements, and ions tracing warm gas tend to show more pronounced evolution than those tracing cold or cool gas. However, the statistical significance of these trends remains limited, and future studies with larger samples are needed to further verify these conclusions.

\textbf{Multi-phase gas distribution:}
The absorption strength decreases with impact parameter, following the relation $W \propto D^{-0.50 \pm 0.38}$ (Figure \ref{fig:D} and Figure \ref{fig:beta}). All gas components can be effectively described by a two-halo model. The radial profiles of low-ionization species (e.g., Mg\,\textsc{ii}, Fe\,\textsc{ii}) exhibit slopes similar to those of galaxy two-point correlation functions at comparable redshifts. In contrast, high-ionization species demonstrate more extended distributions, which aligns with both simulations and previous observational data. 

%Meanwhile, we find that the projected correlation function of QSOs and cold gas exhibits a similar slope to the galaxy two-point correlation function at the same $D$ scale, suggesting that cold gas may play a distinctive role in shaping environments conducive to galaxy formation.

\textbf{Ion-Specific Trends:}
Fe/Mg abundance ratios, gauged by the ratio $W_{\text{FeII}}/W_{\text{MgII}}$, decline with increasing redshift but show little sensitivity to distance (see Figure \ref{fig:femg}). This trend can be attributed to the vigorous star formation activities at high redshift, along with the delayed  Fe enrichment from SNIa in comparison to the Mg production by CCSN. %This trend was also observed by \cite{dey2015}, and is consistent with the predictions of SNIa/CCSN rate evolution by \cite{maoz12} and \cite{madau14}.  
Alternatively, this decline may also indicate a top-heavy IMF in the early universe, which would enhance the production of $\alpha$-elements are produced.

While this study approximates gas surrounding QSOs as an isotropic shell, we recognize that the true structure of the CGM and IGM is inherently patchy and inhomogeneous \citep{lofthouse23,faucher2023}. This complexity underscores the need for enhanced observational capabilities to accurately resolve ionization gradients and their dependencies on host galaxy properties. The release of the DESI Year~1 dataset will provide unprecedented spectral coverage and statistical depth, enabling transformative studies of the distribution and evolution of multiphase gas. Notably, absorption-line spectroscopy, combined with this extensive spectral dataset, offers unique insights into faint, diffuse baryons, particularly in low-density environments like the outskirts of galaxy halos and the cosmic web. By leveraging these advancements, future investigations will be positioned to unveil the multiphase nature of gas accretion and ionization states within dark matter potential wells, bridging the gap between small-scale galaxy feedback and the large-scale baryon cycle.

%This study treats the gas surrounding as an isotropic shell. In fact, the structure of the CGM (and IGM) is much more complex, with patchy and inhomogeneous characteristics \citep{lofthouse23}. Nonetheless, it is promising that the imminent DESI Y1 data release (March 2025) will enhance statistical power for resolving ionization gradients and host galaxy dependencies. At that time, we will be able to conduct more detailed and in-depth studies of the distribution and evolution of  the multiphase gas using absorption line spectroscopy, taking advantage of the extensive data released. \textbf{We emphasize the unique capability of absorption-line spectroscopy to unveil faint, diffuse baryons. The upcoming DESI Year 1 data release (2025), with its unprecedented spectral coverage and statistical power, will transform our ability to characterize the multiphase nature of extremely low-density gas in the outskirts of galaxy halos and along cosmic filaments. This paves the way for uncovering the histories of gas accretion and reionization within dark matter potential wells.}

\section*{Acknowledgements}
The authors especially thank the referee for the insightful comments and suggestions, which are crucial to improving this work. EW thanks the support of the National Science Foundation of China (Nos. 12473008) and the Start-up Fund of the University of Science and Technology of China (No. KY2030000200).
HZ acknowledges the supports from National Key R\&D Program of China (grant Nos. 2022YFA1602902, 2023YFA1607800, 2023YFA1607804) and the National Natural Science Foundation of China (NSFC; grant Nos. 12120101003 and 12373010). 
ZCH is supported by NSFC-11903031 and USTC Research Funds of
the Double First-Class Initiative YD 3440002001.
HYW is supported by the National Natural Science Foundation of China (Nos. 12192224) and CAS Project for Young Scientists in Basic Research, Grant No. YSBR-062. 
YG receives funding from Scientific Research Fund of Dezhou University (3012304024), Shandong Provincial Natural Science Foundation (ZR2024QA212), and the National Natural Science Foundation of China (NSFC, Nos.12033004 and 12233005).
The authors gratefully acknowledge the support of Cyrus Chun Ying Tang Foundations.

\appendix
\section{Comparison with ELGs and LRGs} \label{sec:appenA}
\renewcommand\thefigure{A}
\begin{figure*}
    \centering
    \includegraphics[width = 2\columnwidth]{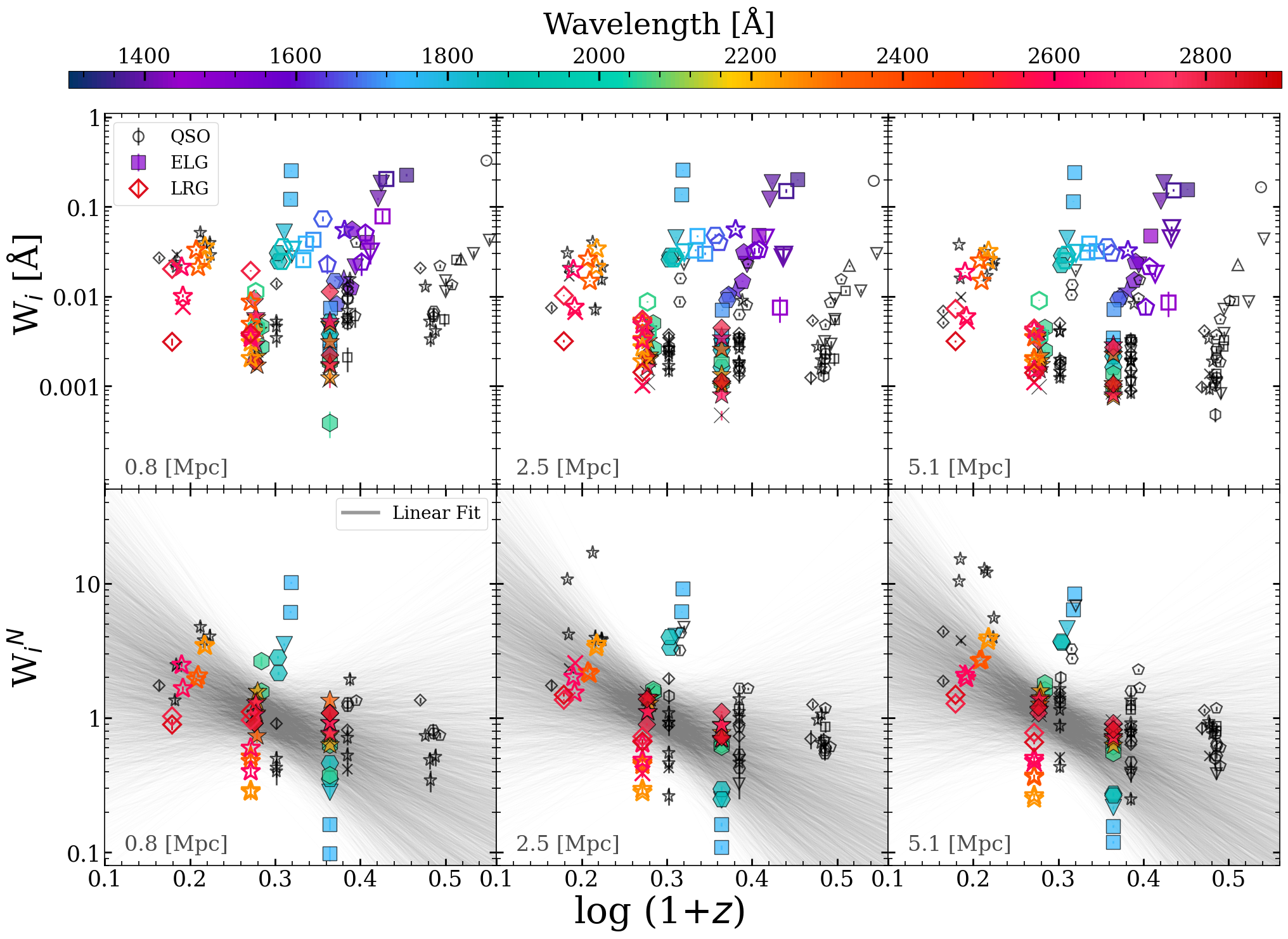}
    \caption{Similar to Figure \ref{fig:metalev}, this plot includes data for ELGs, LRGs, and QSOs. Filled colored markers represent ELGs, hollow colored markers represent LRGs, and hollow black markers represent QSOs.}
    \label{fig:elglrg}
\end{figure*}

To determine whether the evolution of $W$ is a general trend, we also retrieved data for ELGs and LRGs. %(in Table \ref{tab2}). 
Figure \ref{fig:elglrg} presents a trend comparable to that shown in  Figure \ref{fig:metalev}, incorporating data from ELGs and LRGs. The plot demonstrates that \( W_i \) decrease with increasing redshift, as indicated by the linear fits across all three galaxy types. This consistent trend suggests that the evolution of metals in the gas associated with galaxies including QSOs, is cosmic phenomenon rather than being specific to any particular galaxy type.

\section{The impact of mixed stacking} \label{sec:appenC}

As our study involves measuring a large number of absorption lines spanning 1206–2900~\AA, it is indeed unavoidable that some of these features will fall within the Ly$\alpha$ forest region of the background QSO. Therefore, it is important to minimize the confusion in the stacked spectra that could arise from overlap between the regions to the blue and red of Ly$\alpha$. A robust approach, as adopted by \cite{Pieri14} and \cite{Morrison24}, is to stack the spectra within and outside the Ly$\alpha$ forest separately to avoid mixing regions with different physical backgrounds.

The stacked spectra have a high $S/N$; thus, rather than increased noise, we are primarily concerned that mixing may introduce a systematic bias in the $W$ measurements. In Figure \ref{fig:fraction}, we present the PyQSOFit fitting results of background QSO spectra both within and outside the Ly$\alpha$ forest, as well as the fraction of spectra at each rest-frame wavelength that are affected by forest contamination. We also test the equivalent width measurements using only spectra outside the forest ($W_{No-forest}$).  It can be seen that including spectra from the Ly$\alpha$ forest may lead to slightly larger equivalent
widths, but the difference is minimal. And, this difference could be partly due to differences
between the samples. Therefore, such mixed stacking is overall acceptable.

\renewcommand\thefigure{B}
\begin{figure}
    \centering
    \subfigure[The spectrum of the background quasar and its QSOFit, with the region below 1230~Å magnified.]{
        \includegraphics[width=0.45\textwidth]{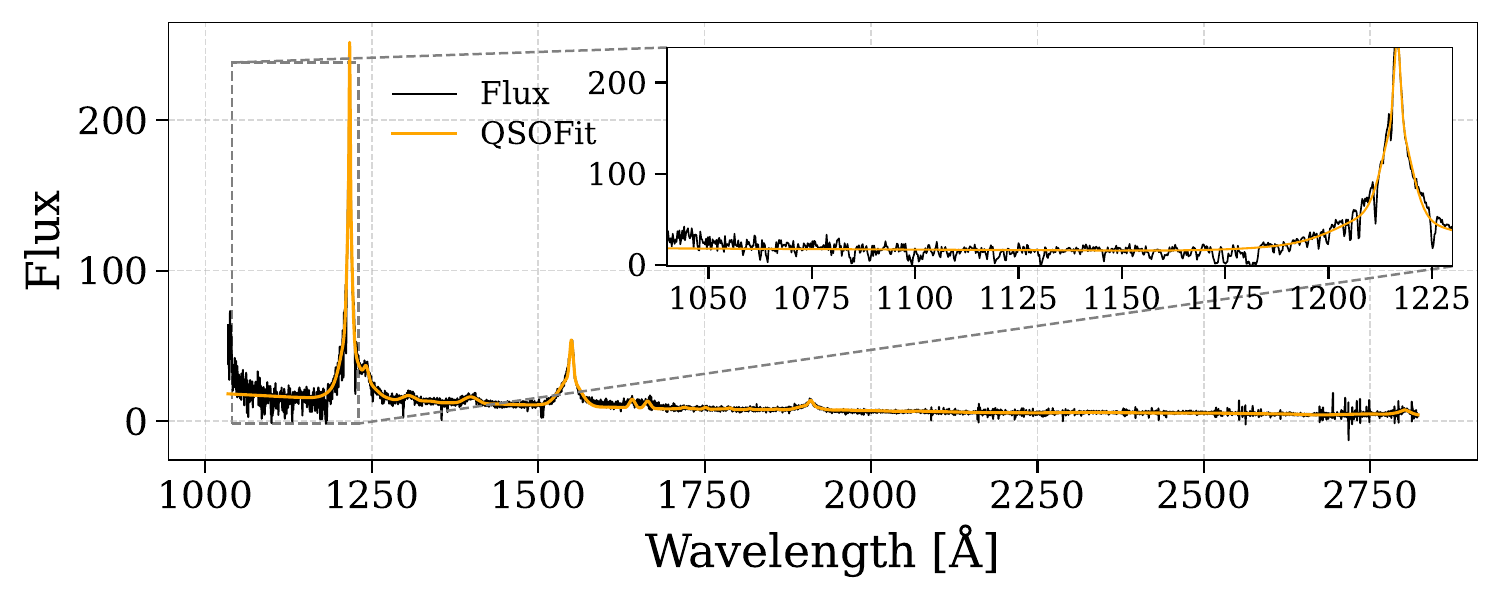}
    }
    \subfigure[The fraction of flux at different wavelengths contributed by the background Ly$\alpha$ forest region. The marker and color represent the element and the absorption line wavelength, respectively, as in Figure \ref{fig:metalev}.
 The inset panel represent the difference in $W$ depending on whether the Ly$\alpha$ absorption forest is excluded.
]{
        \includegraphics[width=0.45\textwidth]{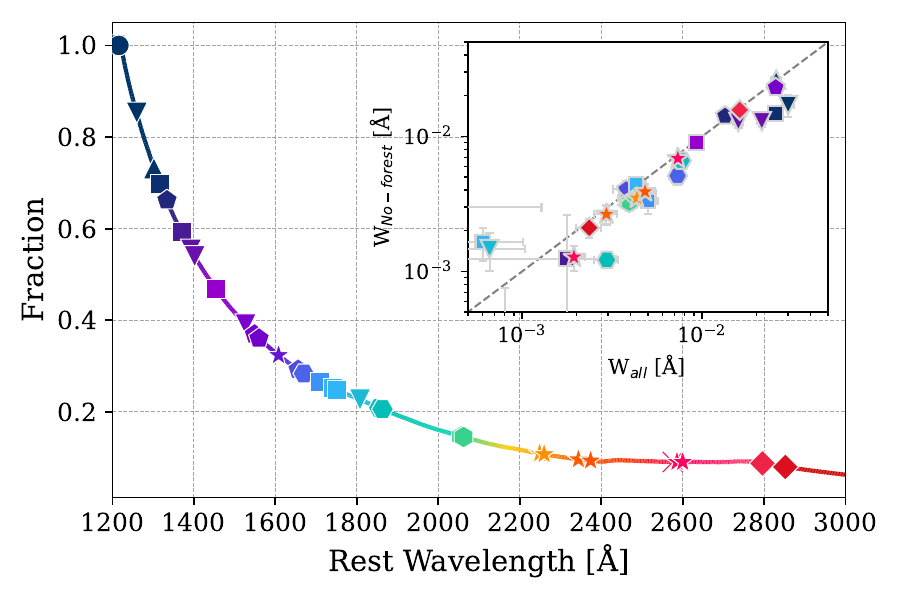}
    }
    \caption{Top panel: Background spectra within and outside the Ly$\alpha$ forest, along with their fits. Bottom panel: For all pairs, the fraction of background spectra falling within the Ly$\alpha$ forest at each foreground rest-frame wavelength, along with the measurement bias in $W$ caused by such mixing.}
    \label{fig:fraction}
\end{figure}

\section{The impact of Energy Level} \label{sec:appenB}

\renewcommand\thefigure{C}
\begin{figure*}
    \centering
    \includegraphics[width=2\columnwidth]{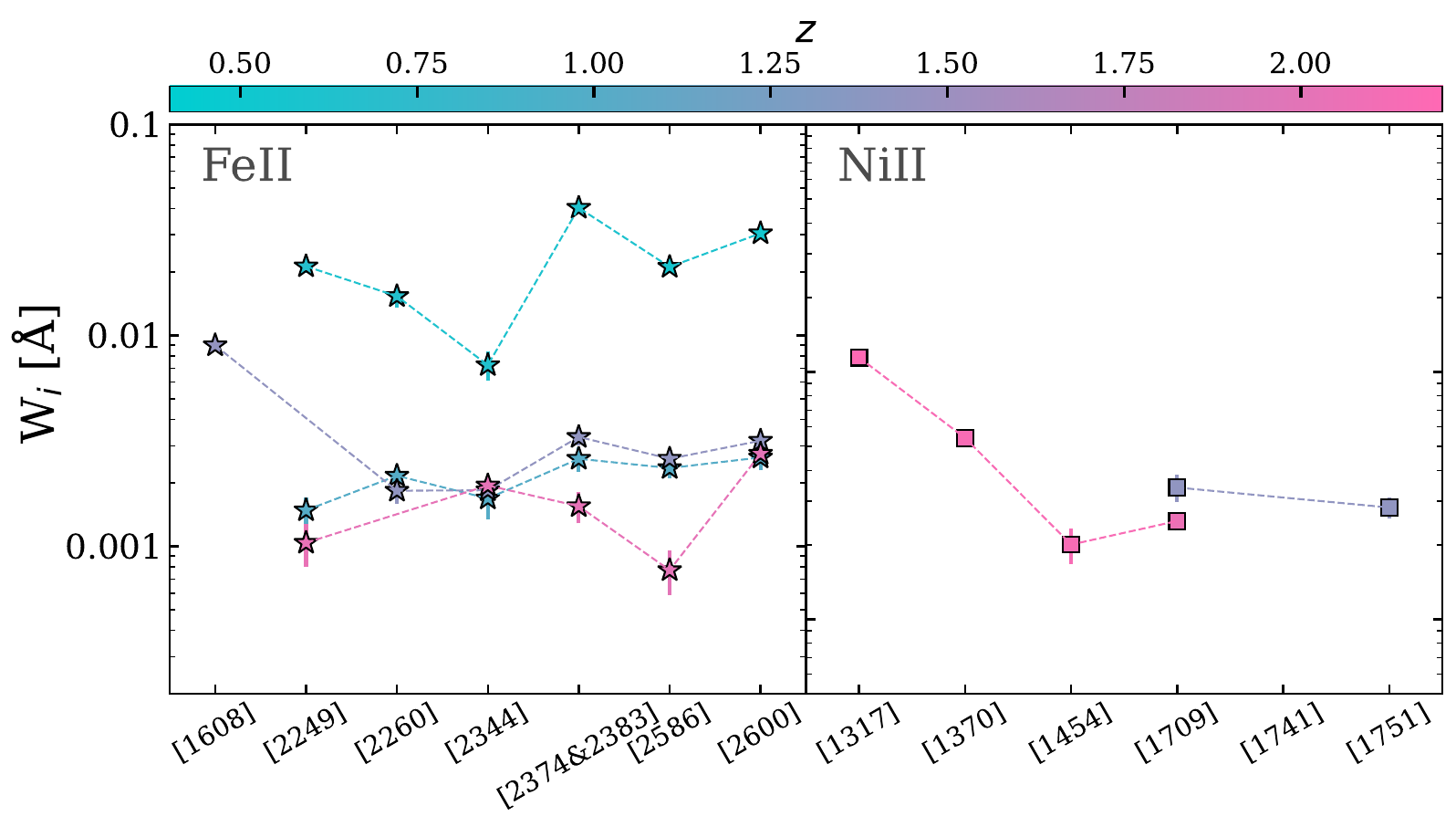}
    \caption{Equivalent width $W_i$ for various absorption lines of Fe\,\textsc{ii} (left panel) and Ni\,\textsc{ii} (right panel). The x-axis represents different absorption lines for each ion, and the color of the data points indicates the redshift \(z\), with the color bar on top showing the corresponding redshift range. 
}
    \label{fig:ion}
\end{figure*}

Figure \ref{fig:ion} presents the equivalent widths \( W \) for different Fe\,\textsc{ii}  (left panel) and Ni\,\textsc{ii} (right panel) absorption lines. The horizontal axis displays different absorption lines for each ion, with Fe \,\textsc{ii} absorption lines on the left and Ni \,\textsc{ii} lines on the right. The figure illustrates that the \( W \) values for Fe \,\textsc{ii} vary significantly with redshift, with trends that are not consistent across all lines. This variability suggests  that different absorption lines corresponding to different physical conditions of the gas. Although the equivalent width is theoretically related to the transition probability of the absorption line, the underlying processes are complex and merit further investigation, especially with the anticipated availability of extensive datasets from future DESI observations.

\bibliography{maix}
\bibliographystyle{aasjournal}

\end{document}